\documentclass[aps,prd,twocolumn,showpacs,preprintnumbers,
               superscriptaddress,longbibliography]{revtex4-2}

\usepackage{amsmath,amssymb,amsfonts}
\usepackage{graphicx}
\usepackage{slashed}
\usepackage{xcolor}

\begin{document}

\title{Pion and kaon electromagnetic form factors in a
flavor-dressed contact interaction: the role of
$N_f$-screening on dynamical chiral symmetry breaking}

\author{\firstname{Aftab}~\surname{Ahmad}}
\email{aftabahmad@gu.edu.pk (corresponding author)}
\affiliation{Institute of Physics, Gomal University,
D.I.~Khan 29220, Khyber Pakhtunkhwa, Pakistan}

\author{\firstname{Muhammad}~\surname{Abid}}
\affiliation{Institute of Physics, Gomal University,
D.I.~Khan 29220, Khyber Pakhtunkhwa, Pakistan}

\author{\firstname{Niaz}~\surname{Hussain}}
\affiliation{Institute of Physics, Gomal University,
D.I.~Khan 29220, Khyber Pakhtunkhwa, Pakistan}

\author{\firstname{Mumtaz}~\surname{Khan}}
\affiliation{Institute of Physics, Gomal University,
D.I.~Khan 29220, Khyber Pakhtunkhwa, Pakistan}

\author{\firstname{Tariq}~\surname{Mahmood}}
\affiliation{Center for High Energy Physics, University of the Punjab,
Lahore 54590, Pakistan}

\author{\firstname{Alfredo}~\surname{Raya}}
\affiliation{Facultad de Ingenier\'ia El\'ectrica,
Universidad Michoacana de San Nicol\'as de Hidalgo,
Edificio Omega, Ciudad Universitaria,
Morelia, Michoac\'an 58030, Mexico}
\affiliation{Centro de Ciencias Exactas,
Universidad del B\'io-B\'io,
Casilla 447, Chill\'an, Chile}

\date{\today}

\begin{abstract}
We compute the electromagnetic form factors of the pion and the
kaon in a symmetry-preserving contact interaction whose effective
coupling depends on the number of active light flavors $N_f$, so that
dynamical chiral symmetry breaking is screened as $N_f$ grows toward the
chiral transition. The calculation uses the rainbow-ladder truncation
of the Schwinger-Dyson and Bethe-Salpeter equations with proper-time
regularization, and it includes the dressed quark-photon vertex obtained
from its own inhomogeneous Bethe-Salpeter equation in the same
truncation. At $N_f=2$ the model reproduces the established contact
interaction values of the meson masses, decay constants and condensates,
and the charge radii $r_\pi = 0.447$~fm and $r_K = 0.417$~fm, which
undershoot experiment by about one third, as is known for this class of
interactions. We therefore restrict every comparison to
$Q^2 \le 0.6$~GeV$^2$, confront it with the NA7 data and with one-loop
chiral perturbation theory, and focus on the response to $N_f$. Both
radii grow monotonically with $N_f$, and the growth survives every
variation of the screening function and of the infrared confinement
scale that we test. The growth is carried by the quark core rather than
by the vector-meson dressing: between $N_f=2$ and $6$ the bare-vertex
contribution to $\langle r_\pi^2\rangle$ rises from $0.097$ to
$0.226$~fm$^2$, while the dressing contribution moves from $0.102$ to
$0.127$~fm$^2$ even though the $\rho$ pole drops from $0.93$ to
$0.70$~GeV. The finite strange mass slows the kaon response only
moderately.
\end{abstract}
\maketitle

\section{Introduction}
\label{sec:intro}

The pion occupies a singular position in QCD. As the lightest
hadron and the pseudo-Goldstone boson of dynamical chiral
symmetry breaking (DCSB), it is both the clearest signal that
chiral symmetry is spontaneously broken in the QCD vacuum and
the simplest bound state available for studying quark
confinement and strong-interaction dynamics. These two roles
are not independent. The pion is light precisely because chiral
symmetry is broken, and its internal structure is shaped by
the same nonperturbative mechanisms that generate the bulk of
visible mass in the universe~\cite{Roberts:2021nhw,
Horn:2016rip}. Any quantitative model of the pion must
therefore capture both faces simultaneously.
Quantum chromodynamics admits two qualitatively distinct
regimes. At short distances the running coupling becomes small,
and the theory is perturbative~\cite{Gross:1973id,
Politzer:1973fx}. At large distances the coupling grows, quarks
and gluons are confined within color-singlet hadrons, and
perturbation theory is useless~\cite{Wilson:1974sk}. The
properties of the pion, including its mass, decay constant,
and electromagnetic structure, are determined entirely in
this nonperturbative sector. Accessing them requires either
lattice QCD or a continuum approach to the Schwinger-Dyson
equations (SDEs) of the theory.

Within the SDE framework, the rainbow-ladder (RL) truncation
provides the minimal consistent scheme that preserves the
axial-vector Ward-Takahashi identity (axWTI) and therefore
correctly encodes Goldstone boson dynamics~\cite{Maris:1997tm,
Munczek:1994zz}. In this setting, the pion emerges as a
simultaneous solution of the quark gap equation and the
homogeneous Bethe-Salpeter equation (BSE). Early studies
using a dressed gluon propagator in the Landau gauge, the
Maris-Tandy model, produced pion and kaon form factors in
good agreement with available data~\cite{Maris:2000sk,
Maris:1999nt}. A simplified but analytically tractable
alternative is the contact interaction (CI) model, in which
the gluon-mediated interaction between quarks is replaced by
a momentum-independent coupling. The CI model, regularized in
a symmetry-preserving way via the Schwinger proper-time scheme,
describes pion static properties well but yields form factors
that are too hard: the charge radii come out too small, and the
form factors approach nonzero constants at large spacelike
momentum transfer, a direct consequence of the
momentum-independent Bethe-Salpeter amplitude~\cite{
GutierrezGuerrero:2010md, Roberts:2010rn, Roberts:2011wy,
Hernandez-Pinto:2023yin}.
It is worth stating plainly where this model sits. A
momentum-independent gluon exchange between quarks is, at the level of
the Lagrangian, a four-fermion contact term, so the framework used here
belongs to the Nambu--Jona-Lasinio (NJL) family~\cite{Nambu:1961tp,
Nambu:1961fr}. Two features separate it from the textbook NJL model.
The first is the proper-time regularization with a second, infrared
cutoff, which removes the real-axis quark production threshold and
thereby supplies a confinement mechanism that NJL lacks. The second is
the explicit dependence of the coupling on the number of active
flavors. Neither modification changes the fact that the rainbow-ladder
truncation of the Schwinger-Dyson and Bethe-Salpeter system is, for a
contact interaction, the same resummation as the Hartree plus
random-phase approximation of the NJL model at leading order in
$1/N_c$. Meson observables in that framework have a long history, and
the pion electromagnetic and scalar form factors in particular have
been computed in the extended NJL model with vector and axial-vector
degrees of freedom~\cite{Bernard:1995hm}; the connection between such
models and chiral effective Lagrangians is reviewed in
Ref.~\cite{Bijnens:1995ww}.

A separate but related question concerns how these properties
change as the number of active light quark flavors $N_f$
varies. QCD with $N_f$ massless quarks in the fundamental
representation of $SU(3)$ loses asymptotic freedom above
$N_f^{AF} = 16.5$~\cite{Politzer:1973fx}. Below this value,
an infrared fixed point of the renormalization group emerges
in a window $N_f^c \lesssim N_f < N_f^{AF}$, commonly called
the conformal window~\cite{Caswell:1974gg, Banks:1981nn,
Appelquist:2009ty}. For $N_f < N_f^c$, the infrared fixed
point shifts to strong coupling, chiral symmetry breaks
dynamically, and quarks confine. At $N_f = N_f^c$, a chiral
phase transition occurs. The precise value of $N_f^c$ is not
yet settled: lattice simulations place it in the range
$8 \lesssim N_f^c \lesssim 12$~\cite{LSD:2014nmn,
Hasenfratz:2016dou, Appelquist:2009ty}, while continuum SDE
studies and effective models suggest $N_f^c \approx 8$~\cite{
bashir2013qcd, Hopfer:2014zna, Ahmad:2020jzn, Ahmad:2022hbu}.
A recent self-consistent SDE analysis including the gluon
propagator finds $N_f^c = 6.81$~\cite{Chen:2026nf}.
The dependence of quark dynamics on $N_f$ has been
investigated in the context of chiral symmetry breaking,
confinement, and the finite-temperature QCD phase diagram
using the CI model with a flavor-dependent effective
coupling~\cite{Ahmad:2020jzn, Ahmad:2022hbu, Ahmad:2023mqg}.
In this framework, the effective coupling
carries an explicit factor of $\sqrt{1-(N_f-2)/\mathcal{N}_f^c}$,
which suppresses quark-gluon interactions as $N_f$ grows and
drives the system toward chiral symmetry restoration. Previous
work within this model established the $N_f$- and $N_c$-dependence
of the dynamical quark mass, the quark-antiquark
condensate~\cite{Ahmad:2020jzn, Ahmad:2022hbu,Ahmad_2026},
and the leptonic decay constants of the pion and
kaon~\cite{Ahmad:2024emu}. What has not been
addressed is how the electromagnetic structure of the pion
and the kaon responds to the screening of the effective coupling
as $N_f$ increases.

This is the question we take up here. Working within the
rainbow-ladder truncation of the SDE-BSE system and employing
the symmetry-preserving Schwinger proper-time regularization
scheme, we compute the electromagnetic form factors of the
charged pion and kaon for varying $N_f$ at fixed $N_c = 3$. The
calculation proceeds in the generalized impulse approximation
following Refs.~\cite{GutierrezGuerrero:2010md, Roberts:2011wy,
Maris:2000sk}, and the photon couples to the quarks through the
dressed quark-photon vertex, which in this truncation obeys its own
inhomogeneous Bethe-Salpeter equation and carries the $\rho$-meson pole.
The kaon provides a complementary window, because its strange
constituent responds to the screening differently from the light one.

The comparison with experiment is confined to the region where a
contact interaction can be read at all. The low-momentum spacelike
pion form factor has been mapped by pion-electron elastic scattering,
most precisely by the CERN NA7 experiment~\cite{Amendolia:1984nz,
Amendolia:1986wj}, with earlier Fermilab measurements by Dally
\emph{et al.}~\cite{Dally:1982zk}; the kaon charge radius was measured
in the same way~\cite{Dally:1980dj, Amendolia:1986ui}. We add one-loop
chiral perturbation theory ($\chi$PT)~\cite{Gasser:1983yg,
Gasser:1984gg, Gasser:1984ux} as a model-independent benchmark of the
low-energy behavior. The momentum transfers reached at Jefferson Lab
and targeted by the Electron-Ion Collider~\cite{Huber:2008id,
Horn:2006tm, Aguilar:2019teb, AbdulKhalek:2021gbh} lie outside that
region, and we make no statement about them.

Two results anticipate the rest of the paper. The model reproduces
the established static properties at $N_f = 2$ but undershoots the
measured charge radii by about one third; this is a known feature of
the contact interaction~\cite{Roberts:2011wy, Chen:2012txa,
Hernandez-Pinto:2023yin}, and it means that the absolute radii are not
predictions to be taken at face value. What the model does control is
the response to $N_f$. We find that both radii grow monotonically as the
coupling is screened, that the growth is robust under every variation of
the two least constrained ingredients of the model, and that it is
driven by the quark core of the bound state rather than by the
vector-meson dressing of the photon vertex.

The rest of this paper is organized as follows. Section~\ref{sec:model}
presents the flavor-dressed contact interaction model, the gap
equation, the physical basis of the $N_f$-dependent confinement scale
and the sensitivity of the results to the model assumptions.
Section~\ref{sec:bse} develops the Bethe-Salpeter formalism, the
static observables, the form factors and the dressed quark-photon
vertex. Section~\ref{sec:results} presents the numerical results, and
Section~\ref{sec:conclusions} offers the conclusions.
Appendix~\ref{app:triangle} collects the expressions for the
triangle diagram with unequal masses.

\section{The Flavor-Dressed Contact Interaction Model}
\label{sec:model}

\subsection{Schwinger-Dyson equation and the dressed quark propagator}

The starting point is the SDE for the
dressed-quark propagator~\cite{Schwinger:1951nm, Roberts:1994hh}.
In Euclidean space and Landau gauge, the full propagator
$S_f(p)$ of a quark with flavor $f$ can be written as
\begin{equation}
S_f^{-1}(p) = S_{f,0}^{-1}(p) + \Sigma_f(p)\,,
\label{eq:SDE}
\end{equation}
where $S_{f,0}(p) = (i\slashed{p} + m_f)^{-1}$ is the bare
propagator, $m_f$ is the current quark mass, and the
self-energy reads
\begin{equation}
\Sigma_f(p) = \int \frac{d^4q}{(2\pi)^4}\, g^2 D_{\mu\nu}(p-q)
\frac{\lambda^a}{2}\gamma_\mu\, S_f(q)\,
\frac{\lambda^a}{2}\Gamma_\nu(p,q)\,.
\label{eq:selfenergy}
\end{equation}
Here $g^2 D_{\mu\nu}$ is the dressed gluon propagator,
$\Gamma_\nu$ the dressed quark-gluon vertex, and
$\lambda^a$ the Gell-Mann matrices, which satisfy
\begin{equation}
\sum_{a=1}^{N_c^2-1}\frac{\lambda^a}{2}\frac{\lambda^a}{2}
= \frac{1}{2}\!\left(N_c - \frac{1}{N_c}\right)\!I \equiv C_F\, I\,,
\label{eq:casimir}
\end{equation}
with $I$ the identity matrix in color space.

\subsection{The flavor-dependent contact interaction}

The flavor-dressed contact interaction (FCI) model replaces
the momentum-dependent gluon exchange with a
momentum-independent effective coupling whose strength depends
explicitly on $N_f$~\cite{Ahmad:2020jzn, Ahmad:2022hbu}.
In the Landau gauge and within the infrared region where
gluons acquire a dynamically generated mass $m_g$~\cite{
Cornwall:1981zr, Aguilar:2015bud, GutierrezGuerrero:2010md},
the model is defined by
\begin{equation}
g^2 D_{\mu\nu}(p-q) = \delta_{\mu\nu}\,\alpha_{\rm eff}(N_f)\,,
\label{eq:FCI}
\end{equation}
where
\begin{equation}
\alpha_{\rm eff}(N_f) = \frac{4\pi\alpha_{\rm ir}}{m_g^2}
\left[1 - \frac{N_f - 2}{\mathcal{N}_f^c}\right]^{p}\,,
\qquad p = \frac{1}{2}\,.
\label{eq:alphaeff}
\end{equation}
Here $\mathcal{N}_f^c$ is \emph{not} the critical flavor number
itself but a shifted version of it, $\mathcal{N}_f^c = N_f^c + \eta$.
The parameter $\alpha_{\rm ir} = 0.93\pi$ sets the
infrared-enhanced interaction strength, and $m_g = 0.8$~GeV
is the gluon mass scale~\cite{Boucaud:2011ug}. The
$N_f$-dependent factor encodes the progressive weakening of the
effective coupling as the number of light quark flavors increases.
The shift $\eta$, in the range $1.8$--$2.3$, is
adjusted to place the onset of chiral symmetry restoration
at $N_f^c \approx 8$, consistent with lattice and continuum
QCD studies~\cite{LSD:2014nmn, bashir2013qcd,
Hopfer:2014zna}. The factor $(N_f - 2)$ in
Eq.~(\ref{eq:alphaeff}) anchors the model at $N_f = 2$, where the
standard CI model is recovered; the physical motivation
for this parametrization is discussed in
Ref.~\cite{Ahmad:2020jzn}. The exponent $p$ is written explicitly
because Sec.~\ref{sec:sensitivity} varies it.

When this model is inserted into the SDE, the quark wave
function renormalization reduces to unity and the dressed
propagator takes the simple form $S_f(p) =
(i\slashed{p} + M_f)^{-1}$, with $M_f$ a
momentum-independent dynamical mass. This
momentum independence is the defining feature of the CI
class of models and the source of their analytical
tractability~\cite{GutierrezGuerrero:2010md, Roberts:2010rn,
Roberts:2011cf}.
All the color and coupling factors enter every equation below
through a single combination,
\begin{equation}
\mathcal{G}(N_f) \equiv \frac{C_F\,\alpha_{\rm eff}(N_f)}{4\pi^2}
= \frac{4\alpha_{\rm ir}}{3\pi m_g^2}
\left[1 - \frac{N_f - 2}{\mathcal{N}_f^c}\right]^{p},
\label{eq:Gcal}
\end{equation}
where the second form holds for $N_c = 3$. At $N_f = 2$ one has
$\mathcal{G} = 1.9375$~GeV$^{-2}$.

\subsection{Gap equation and confinement}

Inserting the FCI model into Eq.~(\ref{eq:SDE}) and
performing the Dirac traces in the rainbow-ladder
truncation, where the dressed vertex is replaced by its
bare counterpart $\Gamma_\nu = \gamma_\nu$, the dynamical
quark mass satisfies
\begin{equation}
M_f = m_f + M_f\,\mathcal{G}(N_f)
\int_0^\infty ds\, \frac{s}{s + M_f^2}\,.
\label{eq:gap_raw}
\end{equation}
The integral is ultraviolet divergent and must be regulated.
The Schwinger proper-time scheme~\cite{Schwinger:1951nm}
introduces two cutoff scales, an ultraviolet regulator
$\tau_{\rm uv} = \Lambda_{\rm uv}^{-1}$ and an infrared regulator
$\tau_{\rm ir} = \Lambda_{\rm ir}^{-1}$,
\begin{eqnarray}
\frac{1}{s + M_f^2} &\to&
\int_{\tau_{\rm uv}^2}^{\tau_{\rm ir}^2} d\tau\,
e^{-\tau(s + M_f^2)} \nonumber \\
&=&
\frac{e^{-\tau_{\rm uv}^2(s+M_f^2)} -
      e^{-\tau_{\rm ir}^2(s+M_f^2)}}{s + M_f^2}\,.
\label{eq:PTreg}
\end{eqnarray}
The role of each regulator is distinct. The ultraviolet
cutoff $\tau_{\rm uv}$ removes quadratic and logarithmic
divergences and sets the scale for all dimensioned
quantities in the model; it cannot be eliminated since the
CI model is non-renormalizable. The infrared cutoff
$\tau_{\rm ir}$, on the other hand, implements quark
confinement: its nonzero value removes the
real-axis quark production threshold from the propagator,
so that the quark never reaches its mass shell and therefore
cannot appear in the physical spectrum~\cite{Ebert:1996vx,
Roberts:2007jh, Roberts:2011cf, Roberts:2011wy}, (see also a recent comparative study on confinement order parameters in~\cite{Ahmad:2026puc}) .

After the integration over $s$, the gap equation
reduces to
\begin{equation}
M_f = m_f + M_f\,\mathcal{G}(N_f)\,
\mathcal{C}(M_f^2)\,,
\label{eq:gap}
\end{equation}
where, suppressing the dependence on the regulators,
\begin{eqnarray}
\mathcal{C}(z) &=& z\left[\Gamma(-1,z\tau_{\rm uv}^2)
- \Gamma(-1,z\tau_{\rm ir}^2)\right], \nonumber\\
\bar{\mathcal{C}}_1(z) &=& -\frac{d\,\mathcal{C}(z)}{dz}
= \Gamma(0,z\tau_{\rm uv}^2) - \Gamma(0,z\tau_{\rm ir}^2),
\label{eq:Cfunctions}\\
\bar{\mathcal{C}}_2(z) &=& -\frac{1}{2}\frac{d\,\bar{\mathcal{C}}_1(z)}{dz}
= \frac{e^{-z\tau_{\rm uv}^2} - e^{-z\tau_{\rm ir}^2}}{2z}, \nonumber
\end{eqnarray}
and $\mathcal{C}_1(z) = z\,\bar{\mathcal{C}}_1(z)$, with
$\Gamma(a,y) = \int_y^\infty t^{a-1}e^{-t}\,dt$ the upper incomplete
Gamma function. Every regularized integral in this paper reduces to one
of these three functions: $\mathcal{C}$ for one propagator,
$\bar{\mathcal{C}}_1$ for two and $\bar{\mathcal{C}}_2$ for three, each
evaluated at the appropriate Feynman-parametrized argument.

When $N_f$ differs from the reference value $N_f = 2$, the infrared
cutoff is adjusted to maintain the connection between DCSB and
confinement through the rescaled confinement scale
\begin{equation}
\tilde{\tau}_{\rm ir} = \tau_{\rm ir}\,
\frac{M_{u}(N_f=2)}{M_{u}(N_f)}\,,
\label{eq:tauir_rescaled}
\end{equation}
For each $N_f$ we iterate the light-quark gap equation and 
Eq.~(\ref{eq:tauir_rescaled}) until $\tilde\tau_{\rm ir}$ and 
$M_u(N_f)$ are mutually consistent, and use the resulting 
$\tilde\tau_{\rm ir}$ for both flavors. The light quark sets the scale 
because it carries the chiral order parameter; the strange mass 
contains a large explicit piece and would tie the confinement scale to 
the current mass rather than to DCSB. The converged infrared scale 
$\tilde\Lambda_{\rm ir} = 1/\tilde\tau_{\rm ir}$ is $0.240$, $0.204$, 
$0.152$ and $0.071$~GeV at $N_f = 2$, $4$, $6$ and $8$. The iteration 
converges quickly, and a single update starting from 
$\tau_{\rm ir}$ already reproduces the converged masses to better 
than $10^{-3}$~GeV.
The light-quark condensate follows directly from the gap equation.
The regularized trace of the propagator gives
\begin{equation}
-\langle\bar{q}q\rangle_f = \frac{N_c}{4\pi^2}\,M_f\,\mathcal{C}(M_f^2)
= \frac{N_c}{4\pi^2}\,\frac{M_f - m_f}{\mathcal{G}(N_f)}\,,
\label{eq:condensate}
\end{equation}
which is the chiral condensate in the chiral limit.

The structure encoded in Eqs.~(\ref{eq:gap})--(\ref{eq:condensate})
has a transparent physical content. As $N_f$ grows,
$\mathcal{G}(N_f)$ falls, the dynamical part of the mass shrinks, and in
the chiral limit the gap equation eventually admits only the trivial
solution. Simultaneously, Eq.~(\ref{eq:tauir_rescaled})
drives $\tilde\tau_{\rm ir}$ upward, the infrared cutoff
$\tilde\Lambda_{\rm ir}$ falls from $0.240$~GeV at $N_f = 2$ to
$0.071$~GeV at $N_f = 8$, and the region of the propagator that
confinement removes shrinks with it.

\subsection{Why the confinement scale follows the chiral order
parameter along $N_f$}
\label{sec:whylink}

Confinement and DCSB are distinct phenomena. At zero temperature
and density they are governed by different scales, and there are
settings in which they separate: at large baryon chemical potential a
confined but chirally symmetric phase has been proposed~\cite{
McLerran:2007qj}, and in two dimensions confinement is linear and
unconditional. Eq.~(\ref{eq:tauir_rescaled}) is therefore a modeling
choice and needs a justification that does not rest on
finite-temperature thermodynamics, where the two transitions happen to
coincide in the crossover region.

The justification we offer is specific to the direction in theory
space that this paper explores. The variable is $N_f$ at $T = 0$ and
$\mu = 0$, and the endpoint is the lower edge of the conformal
window~\cite{Caswell:1974gg, Banks:1981nn, Appelquist:2009ty}. Inside
the window the infrared dynamics is controlled by a fixed point of the
renormalization group, so the theory is scale invariant at long
distances. A scale-invariant theory has neither a string tension nor a
dynamically generated mass. Entering the conformal window from below
therefore removes confinement and DCSB together, and it does so for a
reason that is not thermal: both require a scale, and the fixed point
removes it. Equation~(\ref{eq:tauir_rescaled}) is the simplest
prescription that builds this endpoint into a model with a single
dynamical scale. It does not claim that the two phenomena are the same,
nor that they vanish together anywhere else in the phase diagram.

The prescription can still bias the results, and the test is to
remove it. Section~\ref{sec:sensitivity} repeats the calculation with
$\tau_{\rm ir}$ held at its $N_f = 2$ value, which decouples
confinement from the chiral order parameter completely. The answer is
quantitative and is given in Table~\ref{tab:sens}: the pion radius at
$N_f = 6$ changes from $0.594$ to $0.532$~fm, about $10\%$, and
neither the monotonic growth with $N_f$ nor the ordering
$r_K < r_\pi$ is affected.

\subsection{Sensitivity to the screening function and to the
rescaled confinement scale}
\label{sec:sensitivity}

Equations~(\ref{eq:alphaeff}) and (\ref{eq:tauir_rescaled}) are
the two least constrained ingredients of the model. Neither
is derived from QCD. The square-root screening factor is the simplest
function that is unity at $N_f = 2$, decreases monotonically, and
vanishes at $N_f = \mathcal{N}_f^c + 2$; the rescaling of
$\tau_{\rm ir}$ was discussed above. We repeat the entire
calculation with three families of alternatives: the exponent
$p \in \{0.25, 0.5, 1\}$ in Eq.~(\ref{eq:alphaeff}), which keeps the
endpoint fixed while changing how fast it is approached; the shift
$\eta$ at the two ends of its interval, $1.8$ and $2.3$; and a fixed
$\tau_{\rm ir}$. The results are collected in Table~\ref{tab:sens}.

\begin{table}[t]
\centering
\caption{Sensitivity of the main observables to the model
assumptions, at $N_f = 4$ and $6$. Masses and decay constants in GeV,
radii (dressed vertex) in fm. At $N_f = 2$ all variants coincide with
the first row of Table~\ref{tab:nf}. For $p = 1$ at $N_f = 6$ the pion
is no longer bound.}
\label{tab:sens}
\begin{tabular}{lccccccc}
\hline\hline
Variant & $N_f$ & $M_u$ & $M_s$ & $f_\pi$ & $f_K$ & $r_\pi$ & $r_K$ \\
\hline
baseline & 4 & 0.312 & 0.500 & 0.097 & 0.105 & 0.495 & 0.459 \\
         & 6 & 0.233 & 0.458 & 0.087 & 0.101 & 0.594 & 0.545 \\
$p=0.25$ & 4 & 0.340 & 0.516 & 0.100 & 0.106 & 0.469 & 0.436 \\
         & 6 & 0.303 & 0.495 & 0.097 & 0.105 & 0.503 & 0.466 \\
$p=1$    & 4 & 0.251 & 0.467 & 0.090 & 0.102 & 0.566 & 0.520 \\
         & 6 & 0.079 & 0.390 & -- & -- & -- & -- \\
$\eta=1.8$ & 4 & 0.310 & 0.499 & 0.097 & 0.105 & 0.496 & 0.460 \\
         & 6 & 0.229 & 0.456 & 0.087 & 0.100 & 0.600 & 0.551 \\
$\eta=2.3$ & 4 & 0.314 & 0.501 & 0.098 & 0.105 & 0.493 & 0.457 \\
         & 6 & 0.239 & 0.461 & 0.088 & 0.101 & 0.585 & 0.537 \\
fixed $\tau_{\rm ir}$ & 4 & 0.309 & 0.499 & 0.096 & 0.105 & 0.481 & 0.446 \\
         & 6 & 0.224 & 0.458 & 0.081 & 0.099 & 0.532 & 0.487 \\
\hline\hline
\end{tabular}
\end{table}

Three conclusions follow from the table. The ordering in $N_f$ is
monotonic in every variant, for every observable, and $r_K < r_\pi$
holds throughout. The rate of growth is not universal: between
$N_f = 2$ and $6$ the pion radius grows by $13\%$ for $p = 0.25$ and by
$33\%$ for $p = 0.5$, while for $p = 1$ the pion unbinds before
$N_f = 6$. This is expected, since $p$ controls how fast the
transition is approached, and it means that the numbers quoted at a
given $N_f$ inherit the uncertainty of the screening function.
Changes in $\eta$ move the radii by less than $2\%$, and fixing
$\tau_{\rm ir}$ reduces the growth without reversing it. No qualitative
statement of Sec.~\ref{sec:results} depends on the baseline choices.

\subsection{Model parameters}

The model parameters are fixed at the reference point
$N_f = 2$, $N_c = 3$ by reproducing the
properties of the ground-state $\pi$ and $\rho$ mesons
following Ref.~\cite{Roberts:2011wy}, and the kaon
following Ref.~\cite{Chen:2012qr}. The values are
collected in Table~\ref{tab:params}.

\begin{table}[t]
\centering
\caption{Parameters of the FCI model. All dimensioned
quantities are in GeV. The value of $\mathcal{N}_f^c$
corresponds to $\eta = 1.98$ in $\mathcal{N}_f^c =
N_f^c + \eta$ with $N_f^c = 8$.}
\label{tab:params}
\begin{tabular}{ccccccc}
\hline\hline
$m_{u=d}$ & $m_s$ & $\Lambda_{\rm ir}$ & $\Lambda_{\rm uv}$
& $\alpha_{\rm ir}$ & $m_g$ & $\mathcal{N}_f^c$ \\
\hline
$0.007$ & $0.170$ & $0.240$ & $0.905$ & $0.93\pi$
& $0.800$ & $9.98$ \\
\hline\hline
\end{tabular}
\end{table}

At these parameter values, the solution of the gap and Bethe-Salpeter
equations yields the dynamical masses and meson observables listed in
Table~\ref{tab:results_ref}. They coincide with the published
contact interaction values obtained with the same
parameters~\cite{Roberts:2011wy, Chen:2012qr, Hernandez-Pinto:2023yin},
which we use as a check of our numerical implementation, and they agree
with experiment at the level expected of the truncation.

\begin{table}[t]
\centering
\caption{Static observables at $N_f = 2$, $N_c = 3$, in GeV.
The decay constants are given in the convention
$f_\pi^{\rm exp} \simeq 0.092$~GeV. The experimental values are those
of the Particle Data Group~\cite{ParticleDataGroup:2024cfk}.}
\label{tab:results_ref}
\begin{tabular}{lcccccc}
\hline\hline
 & $m_\pi$ & $m_K$ & $f_\pi$ & $f_K$
& $\kappa_\pi^{1/3}$ & $\kappa_K^{1/3}$ \\
\hline
FCI & $0.140$ & $0.499$ & $0.101$ & $0.106$ & $0.243$ & $0.252$ \\
Exp. & $0.140$ & $0.494$ & $0.092$ & $0.110$ & -- & -- \\
\hline\hline
\end{tabular}
\end{table}

\section{Bethe-Salpeter Equation and Electromagnetic Form Factors}
\label{sec:bse}

\subsection{Pseudoscalar meson bound states}

The pion emerges from the SDE-BSE coupled system as a
simultaneous solution of two equations: the gap equation
already presented in Sec.~\ref{sec:model}, and the
homogeneous BSE for the quark-antiquark bound state.
The BSE determines the meson Bethe-Salpeter amplitude
(BSA) $\Gamma_H(k;P)$ through
\begin{eqnarray}
\left[\Gamma_H(k;P)\right]_{tu} &=& \label{eq:BSE}\\
&&\hspace{-25mm}\int\frac{d^4q}{(2\pi)^4}\,
K_{tu;rs}(k,q;P)\,
\left[S_f(q_+)\Gamma_H(q;P)S_{g}(q_-)\right]_{sr}\,,\nonumber
\end{eqnarray}
where $q_+ = q + \vartheta P$ and
$q_- = q - (1-\vartheta)P$, with
$\vartheta \in [0,1]$ a momentum-sharing parameter that leaves
physical observables unchanged~\cite{Maris:1997tm}.
The indices $r,s,t,u$ collect color, flavor, and spinor
degrees of freedom. The symbol $H = f\bar{g}$ denotes the
flavor content of the meson and $K$ is the quark-antiquark
scattering kernel.

The axWTI imposes
a tight constraint on the BSE kernel~\cite{Maris:1997hd}.
In the chiral limit it reads
\begin{equation}
-P_\mu\Gamma_{5\mu}(k;P) =
S^{-1}(k_+)\gamma_5 + \gamma_5 S^{-1}(k_-)\,,
\label{eq:axWTI}
\end{equation}
with $k_\pm = k \pm P/2$. Any truncation scheme that preserves
this identity guarantees that the pion is a massless
Nambu-Goldstone boson in the chiral limit~\cite{Munczek:1994zz,
Maris:1997hd}.

In the rainbow-ladder truncation combined with the FCI
model, the scattering kernel is
\begin{equation}
K(p,q;P) = -g^2 D_{\mu\nu}(p-q)
\left[\frac{\lambda^a}{2}\gamma_\mu\right]\otimes
\left[\frac{\lambda^a}{2}\gamma_\nu\right],
\label{eq:kernel}
\end{equation}
and the homogeneous BSE for the ground-state pseudoscalar meson
becomes
\begin{eqnarray}
\Gamma_{H}(P) &=& -C_F\,\alpha_{\rm eff}(N_f)\nonumber\\
&&\times\int\frac{d^4q}{(2\pi)^4}\,
\gamma_\mu S_f(q+P)\Gamma_{H}(P)
S_{g}(q)\gamma_\mu\,.
\label{eq:BSE_FCI}
\end{eqnarray}
Within the FCI model the BSA cannot depend on the relative momentum
between the constituents, and the most general decomposition
consistent with Poincar\'e covariance and the quantum numbers of the
meson is~\cite{GutierrezGuerrero:2010md, Roberts:2011wy, Chen:2012qr}
\begin{equation}
\Gamma_{H}(P) =
i\gamma_5 E_{H}(P) +
\frac{1}{2M_R}\gamma_5\gamma{\cdot}P\,F_{H}(P)\,,
\label{eq:BSA}
\end{equation}
where $M_R = M_f M_g/(M_f + M_g)$ is the reduced mass of the
constituents, so that $2M_R = M$ for equal masses. The
pseudovector component $F_H$ is not a correction but an intrinsic
feature of any Poincar\'e-covariant treatment: it materially affects
the form factor and cannot be discarded~\cite{GutierrezGuerrero:2010md,
Maris:1997hd}.
Substituting Eq.~(\ref{eq:BSA}) into Eq.~(\ref{eq:BSE_FCI}),
introducing a Feynman parameter $\alpha$ and regularizing as in
Sec.~\ref{sec:model}, the BSE reduces to the $2 \times 2$ eigenvalue
problem
\begin{equation}
\begin{pmatrix} E_{H} \\ F_{H} \end{pmatrix}
= \mathcal{G}(N_f)
\begin{pmatrix}
\mathcal{K}^{H}_{EE} & \mathcal{K}^{H}_{EF} \\
\mathcal{K}^{H}_{FE} & \mathcal{K}^{H}_{FF}
\end{pmatrix}
\begin{pmatrix} E_{H} \\ F_{H} \end{pmatrix},
\label{eq:BSE_matrix}
\end{equation}
which has a nontrivial solution at $P^2 = -m_{H}^2$.
With $\omega \equiv \omega(\alpha,P^2) = \alpha M_f^2 +
(1-\alpha)M_g^2 + \alpha(1-\alpha)P^2$, the kernels are
\begin{align}
\mathcal{K}^{H}_{EE} &= \int_0^1 \! d\alpha\,
\Big\{\mathcal{C}(\omega) + \big[M_f M_g \nonumber\\
&\quad - \alpha(1-\alpha)P^2 - \omega\big]\bar{\mathcal{C}}_1(\omega)\Big\},\nonumber\\
\mathcal{K}^{H}_{EF} &= \frac{P^2}{2M_fM_g}\int_0^1 d\alpha\,
\big[\alpha M_f^2 + M_fM_g \nonumber\\
&\qquad + (1-\alpha)M_g^2\big]
\bar{\mathcal{C}}_1(\omega), \nonumber\\
\mathcal{K}^{H}_{FE} &= M_R\int_0^1 d\alpha\,
\left[\alpha M_f + (1-\alpha)M_g\right]
\bar{\mathcal{C}}_1(\omega), \nonumber\\
\mathcal{K}^{H}_{FF} &= -\frac{(M_f+M_g)^2}{2M_fM_g}\,
\mathcal{K}^{H}_{FE}\,.
\label{eq:kernels}
\end{align}
For $M_f = M_g$ these reduce to the kernels of
Ref.~\cite{Roberts:2011wy}. The expressions printed in the previous
version of this manuscript for $\mathcal{K}_{FE}$ and
$\mathcal{K}_{FF}$ were dimensionally inconsistent; the ones above were
derived anew and checked numerically against the published
values.

The last line of Eq.~(\ref{eq:kernels}) is where the axWTI enters
the regularization. Preserving Eq.~(\ref{eq:axWTI}) after
regularization requires~\cite{GutierrezGuerrero:2010md, Roberts:2011wy}
\begin{equation}
0 = \int_0^1 d\alpha\left[
\mathcal{C}(\omega) + \mathcal{C}_1(\omega)\right],
\label{eq:axWTI_reg}
\end{equation}
which the proper-time integrals do not satisfy identically. Following
Refs.~\cite{GutierrezGuerrero:2010md, Roberts:2011wy, Chen:2012qr}, it
is imposed as a constraint on the regularized integrals; applied to the
$FF$ channel it produces the relation between $\mathcal{K}_{FF}$ and
$\mathcal{K}_{FE}$ quoted above. With this constraint the pion is
massless in the chiral limit, because at $P^2 = 0$ the eigenvalue
condition coincides with the chiral-limit gap equation.

\subsection{Static observables}

Once the amplitudes $E_{H}$ and $F_{H}$ are
determined, the BSA must be normalized. The canonical
normalization condition in the rainbow-ladder truncation
is~\cite{Maris:1997tm, Roberts:2011wy}
\begin{equation}
P_\mu = N_c\,{\rm tr}_D\!\int\!\frac{d^4q}{(2\pi)^4}\,
\Gamma_{H}(-P)\,\frac{\partial S_f(q+P)}{\partial P_\mu}\,
\Gamma_{H}(P)\, S_{g}(q)\,,
\label{eq:norm}
\end{equation}
evaluated as the derivative of the regularized two-point
function; in the chiral limit it reduces to
$1 = (N_c/4\pi^2)\,\bar{\mathcal{C}}_1(M^2)\,E_\pi(E_\pi - 2F_\pi)$, the
result of Ref.~\cite{Roberts:2011wy}. The leptonic decay constant
and the in-meson condensate are then
\begin{align}
f_{H} &= \frac{N_c}{4\pi^2}\,\frac{1}{M_R}
\left[E_{H}\,\mathcal{K}^{H}_{FE} +
F_{H}\,\mathcal{K}^{H}_{FF}\right],
\label{eq:fdecay}\\
\kappa_{H} &= f_{H}\,\rho_{H}\,,\qquad
\rho_H = \frac{N_c}{4\pi^2}
\left[E_{H}\,\mathcal{K}^{H}_{EE} +
F_{H}\,\mathcal{K}^{H}_{EF}\right],
\label{eq:condensate_meson}
\end{align}
with all kernels at $P^2 = -m_H^2$~\cite{Maris:1997tm, Brodsky:2012ku}.
The decay constant and the in-meson condensate are tied by the
Gell-Mann--Oakes--Renner-type identity
\begin{equation}
f_H\, m_H^2 = (m_f + m_g)\,\rho_H\,,
\label{eq:gmor}
\end{equation}
which a symmetry-preserving truncation must satisfy for any current
masses~\cite{Maris:1997hd}. Since both sides depend on $N_f$ through
different combinations of the same quantities, it is a nontrivial check
of the numerics along the whole flavor trajectory; we return to it in
Sec.~\ref{sec:gmor}.

\subsection{Electromagnetic form factors}

The elastic form factor of the meson $H = f\bar g$ is defined through
the matrix element of the electromagnetic current between one-meson
states. In the generalized impulse approximation it is the sum of the
two diagrams in which the photon couples to the quark or to the
antiquark~\cite{Maris:2000sk, Roberts:1994hh, GutierrezGuerrero:2010md},
\begin{equation}
F_H^{\rm em}(Q^2) = e_f\,F_H^{f}(Q^2) + e_{\bar g}\,F_H^{\bar g}(Q^2)\,,
\label{eq:Fsum}
\end{equation}
with $e_u = 2/3$, $e_{\bar d} = e_{\bar s} = 1/3$, and
\begin{align}
2K_\mu F_H^{f}(Q^2) &=
2N_c\int\frac{d^4t}{(2\pi)^4}\,
{\rm tr}_D\Big[
i\Gamma_H(-p_2)\,S_f(t+p_2)
\nonumber \\
&\quad\times i\Gamma^\gamma_\mu(Q)\,S_f(t+p_1)\,
i\Gamma_H(p_1)\,S_g(t)
\Big],
\label{eq:Fpi_def}
\end{align}
where $p_1 = K - Q/2$ and $p_2 = K + Q/2$ are the
momenta of the initial and final meson, $Q$ is the
spacelike photon momentum, $K$ is the average
meson momentum, and $\Gamma^\gamma_\mu$ is the quark-photon vertex
discussed below. The antiquark contribution $F_H^{\bar g}$ follows from
Eq.~(\ref{eq:Fpi_def}) with the roles of $f$ and $g$ exchanged.

With the bare vertex $\Gamma^\gamma_\mu = \gamma_\mu$, the trace is
reduced in two steps. Numerator terms proportional to the spectator
denominator $t^2 + M_g^2$ are cancelled first, leaving a two-point
integral over the struck-quark propagators; the remainder is a genuine
three-point integral. Both are then Feynman-parametrized and
regularized with the functions of Eq.~(\ref{eq:Cfunctions}). The result
has the structure
\begin{equation}
F_H^{f}(Q^2)\Bigg|_{\rm bare} = E_H^2\,T^f_{EE} + E_HF_H\,T^f_{EF}
+ F_H^2\,T^f_{FF}\,,
\label{eq:Fpi_full}
\end{equation}
with the functions $T^f_{XY}(Q^2)$ given in
Appendix~\ref{app:triangle}. For equal masses they reproduce the
expressions of Refs.~\cite{GutierrezGuerrero:2010md, Roberts:2011wy},
and for unequal masses the $EE$ term coincides with that of
Ref.~\cite{Hernandez-Pinto:2023yin}. In the pion the charge computed with
the canonical normalization of Eq.~(\ref{eq:norm}) is
$F_\pi^{\rm em}(0) = 1.001$, which measures the consistency of the
regularization between the normalization and the triangle. For the
kaon the canonical normalization computed from the light leg and from
the strange leg differ by $4\%$, a known consequence of regularizing
loops with unequal masses; we therefore normalize each flavor
contribution so that $F_H^{f}(0) = F_H^{\bar g}(0) = 1$, which
enforces charge conservation exactly.

\subsection{The dressed quark-photon vertex}
\label{sec:vertex}

In the rainbow-ladder truncation the quark-photon vertex is not an
independent object. It obeys an inhomogeneous Bethe-Salpeter equation
with the same kernel as Eq.~(\ref{eq:kernel}), and for a contact
interaction that equation can be solved in closed
form~\cite{Roberts:2011wy}. The solution is
\begin{equation}
\Gamma^\gamma_\mu(Q) = \gamma_\mu^T\,P_T(Q^2) + \gamma_\mu^L\,,
\qquad
\gamma_\mu^T = \gamma_\mu - \frac{\gamma{\cdot}Q\,Q_\mu}{Q^2}\,,
\label{eq:vertex}
\end{equation}
with $\gamma^L_\mu = \gamma_\mu - \gamma^T_\mu$ undressed, as the vector
Ward-Takahashi identity requires, and
\begin{align}
P_T^f(Q^2) &= \frac{1}{1 + K_\gamma^f(Q^2)}\,, \nonumber\\
K_\gamma^f(Q^2) &= \mathcal{G}(N_f)\int_0^1 d\alpha\,
\alpha(1-\alpha)\,Q^2 \nonumber\\
&\qquad\times\bar{\mathcal{C}}_1\big(M_f^2 + \alpha(1-\alpha)Q^2\big)\,.
\label{eq:PT}
\end{align}
Because $K\cdot Q = 0$ for elastic scattering, only the transverse
part contributes to Eq.~(\ref{eq:Fpi_def}) after contraction with
$K_\mu$, and the dressed form factor is simply
\begin{equation}
F_H^{f}(Q^2) = P_T^f(Q^2)\,F_H^{f}(Q^2)\Bigg|_{\rm bare}\,.
\label{eq:Fdressed}
\end{equation}
The factor satisfies $P_T^f(0) = 1$, so the charge is untouched, and
$P_T^f \to 1$ at large $Q^2$. In between it develops a pole at the
timelike point $Q^2 = -m_\rho^2$ where $1 + K_\gamma^f = 0$, which is the
$\rho$-meson bound-state equation of the model. The vertex is screened
by the same coupling $\mathcal{G}(N_f)$ as everything else, so the
$\rho$ pole moves with $N_f$; we find $m_\rho = 0.929$~GeV at $N_f = 2$,
in agreement with Ref.~\cite{Roberts:2011wy}, decreasing to $0.846$,
$0.704$ and $0.394$~GeV at $N_f = 4$, $6$ and $8$. All form factors and
radii below are computed with the dressed vertex unless stated
otherwise.

\section{Results and Discussion}
\label{sec:results}

\subsection{Dynamical quark masses and static observables}

We solve Eqs.~(\ref{eq:gap}) and (\ref{eq:tauir_rescaled})
self-consistently for the dynamical masses $M_{u}$ and $M_s$ as a
function of $N_f$, at fixed $N_c = 3$, and then the BSE for both
mesons. The results are collected in Table~\ref{tab:nf}.

\begin{table*}[t]
\centering
\caption{Dependence on $N_f$ at $N_c = 3$. Masses and decay
constants in GeV, radii in fm. The labels b and d denote the bare and
the dressed quark-photon vertex, and $m_\rho$ is the pole of
Eq.~(\ref{eq:PT}). The infrared cutoff follows 
Eq.~(\ref{eq:tauir_rescaled}), solved self-consistently.}
\label{tab:nf}
\begin{tabular}{cccccccccccccc}
\hline\hline
$N_f$ & $M_u$ & $M_s$ & $m_\pi$ & $m_K$ & $f_\pi$ & $f_K$
& $\kappa_\pi^{1/3}$ & $\kappa_K^{1/3}$
& $r_\pi^{\rm b}$ & $r_\pi^{\rm d}$ & $r_K^{\rm b}$ & $r_K^{\rm d}$ & $m_\rho$\\
\hline
2 & 0.367 & 0.533 & 0.140 & 0.499 & 0.101 & 0.106 & 0.243 & 0.252 & 0.312 & 0.447 & 0.293 & 0.417 & 0.929\\
4 & 0.312 & 0.500 & 0.141 & 0.500 & 0.097 & 0.105 & 0.238 & 0.250 & 0.366 & 0.495 & 0.341 & 0.459 & 0.846\\
6 & 0.233 & 0.458 & 0.146 & 0.504 & 0.087 & 0.101 & 0.226 & 0.244 & 0.475 & 0.594 & 0.439 & 0.545 & 0.704\\
8 & 0.108 & 0.403 & 0.166 & 0.505 & 0.055 & 0.080 & 0.181 & 0.210 & 0.966 & 1.052 & 0.999 & 1.065 & 0.394\\
\hline\hline
\end{tabular}
\end{table*}

As $N_f$ increases from 2, the coupling $\mathcal{G}(N_f)$
decreases monotonically according to Eq.~(\ref{eq:Gcal}), the strength
of DCSB is reduced, and both dynamical masses fall. The light mass
falls much faster in absolute and relative terms, from $0.367$ to
$0.108$~GeV between $N_f = 2$ and $8$, while the strange mass falls
from $0.533$ to $0.403$~GeV. The reason is that $M_s$ contains a
sizable explicit piece: $m_s = 0.17$~GeV is unaffected by the
screening and acts as a floor below which $M_s$ cannot fall. The
decay constants and the in-meson condensates follow the same trend.
The meson masses first remain nearly constant and then rise near the
end of the trajectory, as the Goldstone mechanism weakens and the
system approaches the symmetric phase, in a pattern analogous to
thermal screening~\cite{Ahmad:2022hbu}.

\subsection{Pion form factor at low momentum transfer}
\label{sec:lowQ2}

\begin{figure}[t]
\centering
\includegraphics[width=0.48\textwidth]{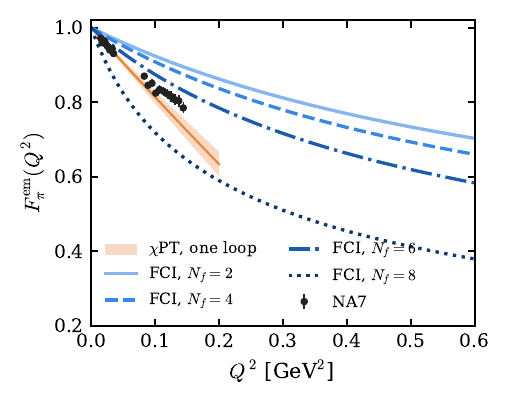}
\caption{Pion electromagnetic form factor, with the dressed
quark-photon vertex, for $N_f = 2, 4, 6, 8$ at $N_c = 3$, compared with
22 of the 45 NA7 data points~\cite{Amendolia:1986wj}, in the 
intervals $0.015 \le Q^2 \le 0.035$~GeV$^2$ and 
$0.083 \le Q^2 \le 0.144$~GeV$^2$ (statistical errors, converted 
from the published $|F_\pi|^2$), and with one-loop $\chi$PT (band, 
$L_9^r = (6.9\pm0.7)\times10^{-3}$), shown only where the latter 
applies.}
\label{fig:Fpi}
\end{figure}

Figure~\ref{fig:Fpi} shows the pion form factor for
$N_f = 2,4,6,8$ in the range $0 \le Q^2 \le 0.6$~GeV$^2$, together
with the NA7 data~\cite{Amendolia:1986wj} and the one-loop $\chi$PT
prediction. For $\chi$PT we use the $SU(3)$ expression of
Refs.~\cite{Gasser:1984gg, Gasser:1984ux},
\begin{eqnarray}
F_{\pi^+}(t) &=& 1 + 2H_{\pi\pi}(t) + H_{KK}(t)\,,\nonumber\\
F_{K^+}(t) &=& 1 + H_{\pi\pi}(t) + 2H_{KK}(t)\,,
\label{eq:chpt}
\end{eqnarray}
with $t = -Q^2$ and
\begin{eqnarray}
H_{PP}(t) &=& \frac{(t - 4M_P^2)\,\bar{J}_P(t)}{12F^2} \nonumber\\
&&+ \frac{t}{F^2}\left[\frac{2L_9^r}{3}
- \frac{\ln(M_P^2/\mu^2) + 1/3}{192\pi^2}\right],
\label{eq:Hfun}
\end{eqnarray}
where $\bar J_P$ is the standard one-loop function~\cite{Gasser:1983yg},
$F = 92.2$~MeV, $\mu = 0.77$~GeV and
$L_9^r(\mu) = (6.9 \pm 0.7)\times 10^{-3}$~\cite{Gasser:1984gg}.
Equation~(\ref{eq:Hfun}) reproduces the one-loop radii
\begin{equation}
\langle r^2_{\pi^+}\rangle = \frac{12L_9^r}{F^2}
- \frac{2\ln(M_\pi^2/\mu^2) + \ln(M_K^2/\mu^2) + 3}{32\pi^2F^2}
\end{equation}
and the corresponding kaon expression, which give $r_\pi = 0.669$~fm and $r_K = 0.641$~fm. The
band reflects the uncertainty in $L_9^r$ and is drawn only up to
$Q^2 = 0.2$~GeV$^2$, beyond which one-loop $\chi$PT is not reliable.

The figure makes the limits of the model visible, and we state them
directly. At $N_f = 2$ the contact interaction gives
$r_\pi = 0.447$~fm against the experimental
$0.659 \pm 0.004$~fm~\cite{ParticleDataGroup:2024cfk}, an undershoot of
about one third, and the curve lies above the data wherever they 
exist; against the 22 NA7 points of Fig.~\ref{fig:Fpi}, with 
statistical errors only, the $\chi^2$ per point is $52$ with the 
dressed vertex and $105$ with the bare one. One-loop $\chi$PT, whose 
only free constant $L_9^r$ was itself fixed from pion data, gives a 
$\chi^2$ per point of $2.0$ below $Q^2 = 0.1$~GeV$^2$ and falls 
below the data above it, where higher orders are needed. The undershoot is the hardness of the contact interaction
form factor~\cite{Roberts:2011wy, Chen:2012txa, Hernandez-Pinto:2023yin},
and it already shows at the level of the slope at the origin. The
model therefore does not predict the absolute pion radius, and we do
not present it as doing so.

What the figure does show reliably is the response to $N_f$. The
form factor softens monotonically as the coupling is screened: at
$Q^2 = 0.3$~GeV$^2$ it falls from $0.812$ at $N_f = 2$ to $0.779$,
$0.716$ and $0.510$ at $N_f = 4$, $6$ and $8$. The radius grows
accordingly (Table~\ref{tab:nf}). This softening has a simple reading
in the model: as $M_u$ decreases, the bound state becomes less
compact, and the scale over which the photon resolves its charge
distribution moves to smaller $Q^2$. That the softening brings the
$N_f = 6$ curve closer to the data is a coincidence of no physical
significance, since the physical world has $N_f = 2+1$ light flavors.

One class of measurements is deliberately absent. The
$e^+e^- \to \pi^+\pi^-$ and $\tau$-decay data~\cite{Barkov:1985ac}
determine the pion form factor in the timelike region, where it is
complex and dominated by the $\rho$ resonance. The quantity computed
here is the spacelike form factor, a real function. A dispersion
relation connects the two, but it requires the two-pion cut, which a
contact interaction does not have, so we do not attempt the
continuation.

\subsection{Kaon form factor and the flavor asymmetry}

\begin{figure}[t]
\centering
\includegraphics[width=0.48\textwidth]{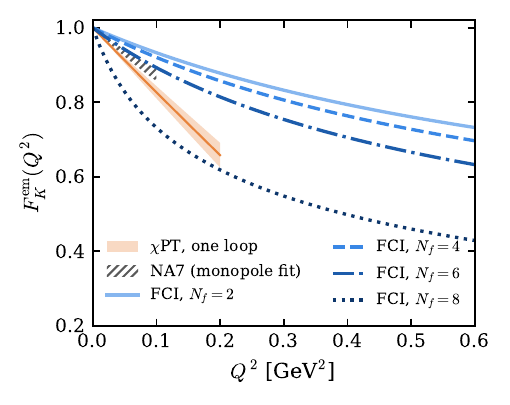}
\caption{Kaon electromagnetic form factor, with the dressed
quark-photon vertex, for $N_f = 2,4,6,8$ at $N_c = 3$. The hatched band
is the NA7 monopole fit with
$\langle r_K^2\rangle = 0.34 \pm 0.05$~fm$^2$~\cite{Amendolia:1986ui},
drawn over the measured range; the orange band is one-loop $\chi$PT.}
\label{fig:FK}
\end{figure}

Figure~\ref{fig:FK} shows the kaon form factor. The pattern is the
same as for the pion, with two differences. The kaon is more compact at
every $N_f \le 6$, $r_K/r_\pi = 0.93$ at $N_f = 2$, in line with
experiment and with $\chi$PT. And its response to the screening is
slower: between $N_f = 2$ and $6$, at $Q^2 = 0.3$~GeV$^2$, the pion form
factor drops by $11.7\%$ and the kaon form factor by $9.4\%$. The
strange current mass, which keeps $M_s$ from falling as fast as $M_u$,
is responsible. The effect is real but moderate. In terms of radii the
two mesons grow by $33\%$ and $31\%$ over the same interval, because
the kaon still contains a light quark, whose dynamical mass is screened
exactly as in the pion.

The ratio $F_K^{\rm em}/F_\pi^{\rm em}$ is larger than unity at
every spacelike $Q^2$ we consider, and it grows with $N_f$: at
$Q^2 = 0.3$~GeV$^2$ it equals $1.025$, $1.034$, $1.052$ and $1.075$ for
$N_f = 2$, $4$, $6$ and $8$. This corrects a statement of the previous
version, which claimed that the ratio approaches unity as $N_f$ grows;
the calculation does not support it in the range where the model can
be used. The radii, in contrast, do converge near the end of the
trajectory, $r_K \simeq r_\pi \simeq 1.06$~fm at $N_f = 8$, where the
pion is barely bound and both mesons are dominated by the soft light
quark. We regard that point as the least reliable of the trajectory and
draw no quantitative conclusion from it.

\subsection{What drives the growth of the radius}
\label{sec:vertexresults}

\begin{figure*}[t]
\centering
\includegraphics[width=0.85\textwidth]{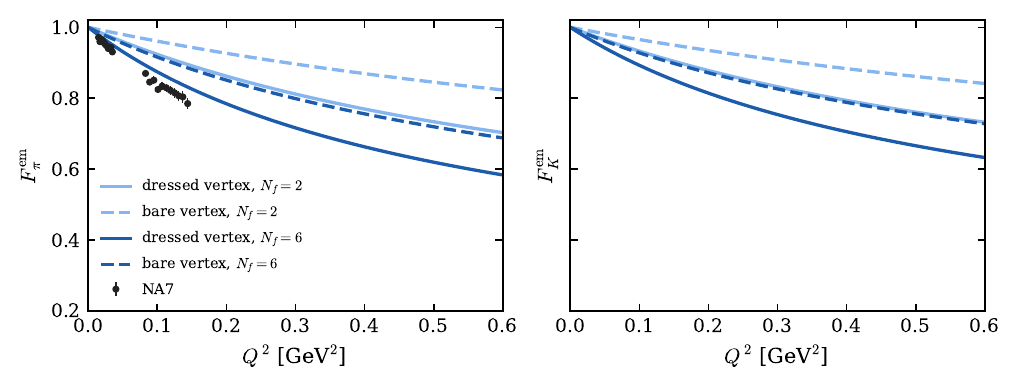}
\caption{Pion (left) and kaon (right) form factors with the bare
(dashed) and the dressed (solid) quark-photon vertex, at $N_f = 2$ and
$N_f = 6$. The NA7 pion data of Fig.~\ref{fig:Fpi} are shown for
reference.}
\label{fig:vertex}
\end{figure*}

\begin{figure}[t]
\centering
\includegraphics[width=0.48\textwidth]{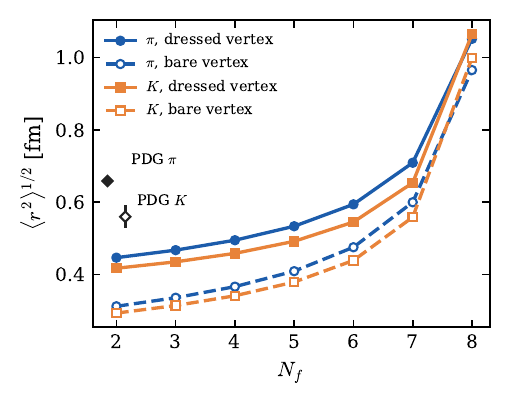}
\caption{Charge radii of the pion and kaon as functions of $N_f$,
with the bare and with the dressed quark-photon vertex. The diamonds
mark the experimental values~\cite{ParticleDataGroup:2024cfk}.}
\label{fig:radii}
\end{figure}

Figure~\ref{fig:vertex} compares the form factors computed with
the bare and with the dressed quark-photon vertex at $N_f = 2$ and $6$,
and Fig.~\ref{fig:radii} shows both radii along the full trajectory.
The dressing enlarges the radius substantially. At $N_f = 2$ it raises
$r_\pi$ from $0.312$ to $0.447$~fm and $r_K$ from $0.293$ to
$0.417$~fm, in agreement with Ref.~\cite{Roberts:2011wy}. It does not
change the sign of any trend, and it does not change the ordering
$r_K < r_\pi$.

The separation into the two pieces is the result we consider most
informative. Since the dressing multiplies the bare form factor, the
mean-square radius splits additively,
$\langle r^2\rangle = \langle r^2\rangle_{\rm bare}
+ 6\,dK_\gamma/dQ^2|_0$. For the pion, between $N_f = 2$ and $6$, the
bare contribution rises from $0.097$ to $0.226$~fm$^2$, while the
dressing contribution moves from $0.102$ to $0.127$~fm$^2$. For the
kaon the corresponding numbers are $0.086 \to 0.193$ and
$0.088 \to 0.105$~fm$^2$. The vector-meson pole drops by about a
quarter over the same interval, so one might have expected the dressing
to dominate the growth. It does not. The growth of the electromagnetic
size under screening is carried by the quark core of the bound state,
which responds directly to the falling dynamical mass, and only
secondarily by the $\rho$-like dressing of the photon.

\subsection{Comparison of the radii with other approaches}

Table~\ref{tab:radii} places the $N_f = 2$ radii in context. Our
values coincide with the earlier contact interaction calculations with
the same parameters, and all of them fall short of experiment and of
$\chi$PT by a similar fraction. The ratio $r_K/r_\pi$ is closer to
experiment than either radius separately: the model gives $0.93$,
$\chi$PT $0.96$, and the PDG averages $0.85 \pm 0.05$. The difference
of mean-square radii, $\langle r_\pi^2\rangle - \langle r_K^2\rangle
= 0.025$~fm$^2$, is smaller than the NA7 value
$0.10 \pm 0.045$~fm$^2$~\cite{Amendolia:1986ui}.

\begin{table}[t]
\centering
\caption{Charge radii (fm) at $N_f = 2$.}
\label{tab:radii}
\begin{tabular}{lcc}
\hline\hline
 & $r_\pi$ & $r_K$ \\
\hline
This work, bare vertex & 0.312 & 0.293 \\
This work, dressed vertex & 0.447 & 0.417 \\
CI, Ref.~\cite{Roberts:2011wy} & 0.45 & -- \\
CI, Ref.~\cite{Hernandez-Pinto:2023yin} & 0.45 & 0.42 \\
One-loop $\chi$PT, Eq.~(\ref{eq:chpt}) & $0.669 \pm 0.029$ & $0.641 \pm 0.030$ \\
NA7~\cite{Amendolia:1986wj, Amendolia:1986ui} & $0.663 \pm 0.006$ & $0.58 \pm 0.04$ \\
Dally \emph{et al.}~\cite{Dally:1980dj} & -- & $0.53 \pm 0.05$ \\
PDG~\cite{ParticleDataGroup:2024cfk} & $0.659 \pm 0.004$ & $0.560 \pm 0.031$ \\
\hline\hline
\end{tabular}
\end{table}

\subsection{The Gell-Mann--Oakes--Renner identity along the
flavor trajectory}
\label{sec:gmor}

Equation~(\ref{eq:gmor}) has no reason to survive the screening
automatically, because $f_H$, $m_H$ and $\rho_H$ each evolve with $N_f$
through different combinations of $E_H$, $F_H$ and the kernels. We find
that it holds to better than one part in $10^4$ for both mesons at every
$N_f$ in Table~\ref{tab:nf}, including $N_f = 8$, where the pion is
barely bound. This confirms that the screened coupling, the rescaled
infrared cutoff and the axWTI constraint of Eq.~(\ref{eq:axWTI_reg})
are mutually consistent, and that the trends reported above are not
artifacts of the numerical implementation. A discussion of the $N_f$
dependence of the decay constants within this model is given in
Ref.~\cite{Ahmad:2024emu}.

\section{Conclusions}
\label{sec:conclusions}

We have studied the electromagnetic form factors of the charged pion
and kaon within a symmetry-preserving, flavor-dressed contact
interaction model of quarks, in which the effective coupling depends
on the number of active light flavors and is progressively screened as
$N_f$ grows toward the chiral transition. The photon couples to
the quarks through the dressed quark-photon vertex obtained in the same
truncation, and the calculation has been confronted with the NA7 data
and with one-loop chiral perturbation theory in the only momentum
range, $Q^2 \le 0.6$~GeV$^2$, where a contact interaction can be read.

At $N_f = 2$ the model reproduces the established contact
interaction values of the masses, decay constants, condensates and
radii, and it inherits their main limitation: the charge radii,
$0.447$~fm for the pion and $0.417$~fm for the kaon, fall short of
experiment by about one third. We do not regard these absolute values
as predictions.

The response to $N_f$ is where the model has something to say.
Both radii grow monotonically as the coupling is screened, from
$0.447$ to $0.594$~fm for the pion and from $0.417$ to $0.545$~fm for
the kaon between $N_f = 2$ and $6$. The growth survives every
variation of the screening function and of the infrared confinement
scale that we tested, including one in which confinement is decoupled
from the chiral order parameter altogether, although its rate depends
on the shape of the screening function. The kaon responds more slowly
than the pion because the strange current mass keeps $M_s$ from
falling as fast as $M_u$, but the difference is moderate, since the
kaon still carries a light quark. The Gell-Mann--Oakes--Renner
identity holds along the whole trajectory.

The most informative outcome is the decomposition of the radius.
The growth of the electromagnetic size under screening comes mainly
from the quark core of the bound state, whose contribution to the pion
mean-square radius more than doubles between $N_f = 2$ and $6$, and
only secondarily from the vector-meson dressing of the photon, whose
contribution grows by a quarter even though the $\rho$ pole drops by
about the same fraction. Within this framework, then, the size of the
lightest pseudoscalars tracks the strength of DCSB through the quark
mass function itself. Testing whether this statement survives a
momentum-dependent interaction, where the absolute radii are
realistic, is the natural continuation of this work.

\begin{acknowledgments}
A.~Raya acknowledges support from SECIHTI (Mexico) under grant
CBF-2026-465 and from UMSNH-CIC under grant 18371.
A.~Ahmad thanks the Institute of Physics, Gomal University for support.
\end{acknowledgments}

\appendix

\section{Triangle diagram with unequal masses}
\label{app:triangle}

For a meson $H = f\bar g$ with $P^2 = -m_H^2$ and the photon
attached to the quark $f$, the bare-vertex form factor of
Eq.~(\ref{eq:Fpi_full}) is, up to a normalization fixed by
$F_H^f(0) = 1$,
\begin{align}
T^f_{XY}(Q^2) &= b_{XY}\int_0^1 d\alpha\,
\bar{\mathcal{C}}_1(\omega_b) \nonumber\\
&\hspace{-6mm}+ 2\int_0^1 du\,u\int_0^1 d\beta
\Big\{a_{XY}\left[\bar{\mathcal{C}}_1(\Delta)
- \Delta\,\bar{\mathcal{C}}_2(\Delta)\right]
+ r_{XY}\,\bar{\mathcal{C}}_2(\Delta)\Big\},
\label{eq:Tapp}
\end{align}
with $\omega_b = M_f^2 + \alpha(1-\alpha)Q^2$ and
\begin{equation}
\Delta = uM_f^2 + (1-u)M_g^2 - u(1-u)m_H^2 
+ u^2\beta(1-\beta)Q^2.
\end{equation}
Writing $S = M_f + M_g$, $D = M_f - M_g$ and
$k = S^2/(4M_f^2M_g^2)$, the coefficients are
\begin{align}
b_{EE} &= 1, \quad a_{EE} = 0, \nonumber\\
r_{EE} &= u\,(D^2 + m_H^2) + 2M_g D, \nonumber\\[3pt]
b_{EF} &= -\frac{S}{M_f}, \quad a_{EF} = -\frac{SD}{2M_fM_g}, \nonumber\\
r_{EF} &= \frac{S}{2M_fM_g}\Big\{Q^2u\,(Du + M_g)
- 4D\,m_H^2\,u(1-u) \nonumber\\
&\qquad\qquad - 2M_g\,(SD + m_H^2)\Big\}, \nonumber\\[3pt]
b_{FF} &= k\,(Q^2 + m_H^2), \quad
a_{FF} = -k\left[u\,(Q^2 + 3m_H^2) - m_H^2\right], \nonumber\\
r_{FF} &= k\Big\{Q^4u^3\beta(1-\beta)
+ Q^2\left[m_H^2u^2(2u-1) - M_gSu\right] \nonumber\\
&\qquad + m_H^2\left[M_f^2u - 2M_fM_gu + 2M_fM_g
- 3M_g^2u + 2M_g^2\right] \nonumber\\
&\qquad + m_H^4\,u\,(2u-1)^2\Big\}.
\label{eq:coeffs}
\end{align}
The antiquark contribution follows by exchanging $M_f$ and $M_g$. For
$M_f = M_g$ these expressions reduce to those of
Refs.~\cite{GutierrezGuerrero:2010md, Roberts:2011wy}, and the $EE$
coefficient coincides with that of Ref.~\cite{Hernandez-Pinto:2023yin}
for unequal masses. We have verified numerically that
Eqs.~(\ref{eq:Tapp}) and (\ref{eq:coeffs}) reproduce the triangle
computed directly from the Dirac trace to machine precision.

\bibliography{sn-bibliography}

@article{Gross:1973id,
    author = "Gross, David J. and Wilczek, Frank",
    editor = "Taylor, J. C.",
    title = "{Ultraviolet Behavior of Nonabelian Gauge Theories}",
    doi = "10.1103/PhysRevLett.30.1343",
    journal = "Phys. Rev. Lett.",
    volume = "30",
    pages = "1343--1346",
    year = "1973"
}

@article{Politzer:1973fx,
    author = "Politzer, H. David",
    editor = "Taylor, J. C.",
    title = "{Reliable Perturbative Results for Strong Interactions?}",
    doi = "10.1103/PhysRevLett.30.1346",
    journal = "Phys. Rev. Lett.",
    volume = "30",
    pages = "1346--1349",
    year = "1973"
}

@article{Wilson:1974sk,
    author = "Wilson, Kenneth G.",
    editor = "Taylor, J. C.",
    title = "{Confinement of Quarks}",
    reportNumber = "CLNS-262",
    doi = "10.1103/PhysRevD.10.2445",
    journal = "Phys. Rev. D",
    volume = "10",
    pages = "2445--2459",
    year = "1974"
}

@article{Appelquist:2009ty,
    author = "Appelquist, Thomas and Fleming, George T. and Neil, Ethan T.",
    title = "{Lattice Study of Conformal Behavior in SU(3) Yang-Mills Theories}",
    eprint = "0901.3766",
    archivePrefix = "arXiv",
    primaryClass = "hep-ph",
    doi = "10.1103/PhysRevD.79.076010",
    journal = "Phys. Rev. D",
    volume = "79",
    pages = "076010",
    year = "2009"

}

@article{LSD:2014nmn,
    author = "Appelquist, T. and others",
    collaboration = "LSD",
    title = "{Lattice simulations with eight flavors of domain wall fermions in SU(3) gauge theory}",
    eprint = "1405.4752",
    archivePrefix = "arXiv",
    primaryClass = "hep-lat",
    reportNumber = "LLNL-JRNL-665913",
    doi = "10.1103/PhysRevD.90.114502",
    journal = "Phys. Rev. D",
    volume = "90",
    number = "11",
    pages = "114502",
    year = "2014"
}

@article{Caswell:1974gg,
    author = "Caswell, William E.",
    title = "{Asymptotic Behavior of Nonabelian Gauge Theories to Two Loop Order}",
    reportNumber = "PRINT-74-1058 (PRINCETON)",
    doi = "10.1103/PhysRevLett.33.244",
    journal = "Phys. Rev. Lett.",
    volume = "33",
    pages = "244",
    year = "1974"
}

@article{Banks:1981nn,
    author = "Banks, Tom and Zaks, A.",
    title = "{On the Phase Structure of Vector-Like Gauge Theories with Massless Fermions}",
    reportNumber = "TAUP-944-81",
    doi = "10.1016/0550-3213(82)90035-9",
    journal = "Nucl. Phys. B",
    volume = "196",
    pages = "189--204",
    year = "1982"
}

@article{Hasenfratz:2016dou,
    author = "Hasenfratz, Anna and Schaich, David",
    title = "{Nonperturbative $\beta$ function of twelve-flavor SU(3) gauge theory}",
    eprint = "1610.10004",
    archivePrefix = "arXiv",
    primaryClass = "hep-lat",
    doi = "10.1007/JHEP02(2018)132",
    journal = "JHEP",
    volume = "02",
    pages = "132",
    year = "2018"
}

@article{bashir2013qcd,
  title={QCD: restoration of chiral symmetry and deconfinement for large N f},
  author={Bashir, A and Raya, A and Rodriguez-Quintero, J},
  journal={Physical Review D},
  volume={88},
  number={5},
  pages={054003},
  year={2013},
  publisher={APS}
}

@article{Hopfer:2014zna,
    author = "Hopfer, Markus and Fischer, Christian S. and Alkofer, Reinhard",
    title = "{Running coupling in the conformal window of large-Nf QCD}",
    eprint = "1405.7031",
    archivePrefix = "arXiv",
    primaryClass = "hep-ph",
    doi = "10.1007/JHEP11(2014)035",
    journal = "JHEP",
    volume = "11",
    pages = "035",
    year = "2014"
}

@article{Ahmad:2020jzn,
    author = "Ahmad, Aftab and Bashir, Adnan and Bedolla, Marco A. and Cobos-Mart\'\i{}nez, J. J.",
    title = "{Flavor, temperature and magnetic field dependence of the QCD phase diagram: magnetic catalysis and its inverse}",
    eprint = "2008.03847",
    archivePrefix = "arXiv",
    primaryClass = "hep-ph",
    doi = "10.1088/1361-6471/abd88f",
    journal = "J. Phys. G",
    volume = "48",
    number = "7",
    pages = "075002",
    year = "2021"
}

@article{Ahmad:2023mqg,
    author = "Ahmad, Aftab and Farooq, Akif",
    title = "{Schwinger Pair Production in QCD from Flavor-Dependent Contact Interaction Model of Quarks}",
    eprint = "2302.13265",
    archivePrefix = "arXiv",
    primaryClass = "hep-ph",
    doi = "10.1007/s13538-024-01581-0",
    journal = "Braz. J. Phys.",
    volume = "54",
    number = "6",
    pages = "212",
    year = "2024"
}

@article{Schwinger:1951nm,
    author = "Schwinger, Julian S.",
    editor = "Milton, K. A.",
    title = "{On gauge invariance and vacuum polarization}",
    doi = "10.1103/PhysRev.82.664",
    journal = "Phys. Rev.",
    volume = "82",
    pages = "664--679",
    year = "1951"
}

@article{Aguilar:2019teb,
    author = "Aguilar, Arlene C. and others",
    title = "{Pion and Kaon Structure at the Electron-Ion Collider}",
    eprint = "1907.08218",
    archivePrefix = "arXiv",
    primaryClass = "nucl-ex",
    reportNumber = "NJU-INP 001/19",
    doi = "10.1140/epja/i2019-12885-0",
    journal = "Eur. Phys. J. A",
    volume = "55",
    number = "10",
    pages = "190",
    year = "2019"
}

@article{Ahmad:2024emu,
    author = "Ahmad, Aftab",
    title = "{$\pi $- and K-Mesons Properties for Large $N_f$}",
    eprint = "2401.11186",
    archivePrefix = "arXiv",
    primaryClass = "hep-ph",
    doi = "10.1007/s10773-026-06298-1",
    journal = "Int. J. Theor. Phys.",
    volume = "65",
    number = "3",
    pages = "88",
    year = "2026"
}

@article{Ahmad_2026,
author={Ahmad, Aftab},
title={QCD phase diagram for large Nf : analysis from contact interaction effective potential},  doi={10.31349/RevMexFis.72.030801},  journal={Rev.Mex. Fís.}, volume={72}, pages={030801–030814},  year={2026}}

@article{Maris:1997hd,
    author = "Maris, Pieter and Roberts, Craig D. and Tandy, Peter C.",
    title = "{Pion mass and decay constant}",
    eprint = "nucl-th/9707003",
    archivePrefix = "arXiv",
    reportNumber = "ANL-PHY-8753-TH-97, KSUCNR-103-97",
    doi = "10.1016/S0370-2693(97)01535-9",
    journal = "Phys. Lett. B",
    volume = "420",
    pages = "267--273",
    year = "1998"
}

@article{Chen:2012qr,
    author = "Chen, Chen and Chang, Lei and Roberts, Craig D. and Wan, Shaolong and Wilson, David J.",
    title = "{Spectrum of hadrons with strangeness}",
    eprint = "1204.2553",
    archivePrefix = "arXiv",
    primaryClass = "nucl-th",
    doi = "10.1007/s00601-012-0466-3",
    journal = "Few Body Syst.",
    volume = "53",
    pages = "293--326",
    year = "2012"
}

@article{Chen:2012txa,
    author = "Chen, Chen and Chang, Lei and Roberts, Craig D. and Schmidt, Sebastian M. and Wan, Shaolong and Wilson, David J.",
    title = "{Features and flaws of a contact interaction treatment of the kaon}",
    eprint = "1212.2212",
    archivePrefix = "arXiv",
    primaryClass = "nucl-th",
    doi = "10.1103/PhysRevC.87.045207",
    journal = "Phys. Rev. C",
    volume = "87",
    pages = "045207",
    year = "2013"
}

@article{Maris:1997tm,
    author = "Maris, Pieter and Roberts, Craig D.",
    title = "{Pi- and K meson Bethe-Salpeter amplitudes}",
    eprint = "nucl-th/9708029",
    archivePrefix = "arXiv",
    reportNumber = "ANL-PHY-8788-TH-97",
    doi = "10.1103/PhysRevC.56.3369",
    journal = "Phys. Rev. C",
    volume = "56",
    pages = "3369--3383",
    year = "1997"
}

@article{Brodsky:2012ku,
    author = "Brodsky, Stanley J. and Roberts, Craig D. and Shrock, Robert and Tandy, Peter C.",
    title = "{Confinement contains condensates}",
    eprint = "1202.2376",
    archivePrefix = "arXiv",
    primaryClass = "nucl-th",
    reportNumber = "SLAC-PUB-14869",
    doi = "10.1103/PhysRevC.85.065202",
    journal = "Phys. Rev. C",
    volume = "85",
    pages = "065202",
    year = "2012"
}

@article{Hernandez-Pinto:2023yin,
    author = "Hern\'andez-Pinto, R. J. and Guti\'errez-Guerrero, L. X. and Bashir, A. and Bedolla, M. A. and Higuera-Angulo, I. M.",
    title = "{Electromagnetic form factors and charge radii of pseudoscalar and scalar mesons: A comprehensive contact interaction analysis}",
    eprint = "2301.11881",
    archivePrefix = "arXiv",
    primaryClass = "hep-ph",
    reportNumber = "JLAB-THY-23-3748",
    doi = "10.1103/PhysRevD.107.054002",
    journal = "Phys. Rev. D",
    volume = "107",
    number = "5",
    pages = "054002",
    year = "2023"
}

@article{Ahmad:2026puc,
    author = "Ahmad, Aftab and Jamil, Muhammad",
    title = "{Confinement order parameters in QCD: a view from Schwinger-Dyson equations and related low energy QCD models}",
    doi = "10.1088/1674-1137/ae8cf3",
    journal = "Chin. Phys. C",
    volume = "50",
    number = "11",
    pages = "113102",
    year = "2026"
}

@article{Ahmad:2022hbu,
    author = "Ahmad, Aftab and Murad, Ali",
    title = "{Color-flavor dependence of the Nambu-Jona-Lasinio model and QCD phase diagram}",
    eprint = "2201.09980",
    archivePrefix = "arXiv",
    primaryClass = "hep-ph",
    doi = "10.1088/1674-1137/ac6cd8",
    journal = "Chin. Phys. C",
    volume = "46",
    number = "8",
    pages = "083109",
    year = "2022"
}

@article{GutierrezGuerrero:2010md,
      author         = "Gutierrez-Guerrero, L. X. and Bashir, A. and Cloet, I. C.
                        and Roberts, C. D.",
      title          = "{Pion form factor from a contact interaction}",
      journal        = "Phys. Rev.",
      volume         = "C81",
      year           = "2010",
      pages          = "065202",
      doi            = "10.1103/PhysRevC.81.065202",
      eprint         = "1002.1968",
      archivePrefix  = "arXiv",
      primaryClass   = "nucl-th",
      SLACcitation   = "%%CITATION = ARXIV:1002.1968;%%"
}

@article{Roberts:2010rn,
      author         = "Roberts, H. L. L. and Roberts, C. D. and Bashir, A. and
                        Gutierrez-Guerrero, L. X. and Tandy, P. C.",
      title          = "{Abelian anomaly and neutral pion production}",
      journal        = "Phys. Rev.",
      volume         = "C82",
      year           = "2010",
      pages          = "065202",
      doi            = "10.1103/PhysRevC.82.065202",
      eprint         = "1009.0067",
      archivePrefix  = "arXiv",
      primaryClass   = "nucl-th",
      SLACcitation   = "%%CITATION = ARXIV:1009.0067;%%"
}

@article{Cornwall:1981zr,
    author = "Cornwall, John M.",
    title = "{Dynamical Mass Generation in Continuum QCD}",
    reportNumber = "UCLA-81-TEP-30",
    doi = "10.1103/PhysRevD.26.1453",
    journal = "Phys. Rev. D",
    volume = "26",
    pages = "1453",
    year = "1982"
}

@article{Aguilar:2015bud,
    author = "Aguilar, A. C. and Binosi, D. and Papavassiliou, J.",
    title = "{The Gluon Mass Generation Mechanism: A Concise Primer}",
    eprint = "1511.08361",
    archivePrefix = "arXiv",
    primaryClass = "hep-ph",
    doi = "10.1007/s11467-015-0517-6",
    journal = "Front. Phys. (Beijing)",
    volume = "11",
    number = "2",
    pages = "111203",
    year = "2016"
}

@article{Roberts:2011wy,
      author         = "Roberts, H. L. L. and Bashir, A. and Gutierrez-Guerrero,
                        L. X. and Roberts, C. D. and Wilson, D. J.",
      title          = "{pi- and rho-mesons, and their diquark partners, from a
                        contact interaction}",
      journal        = "Phys. Rev.",
      volume         = "C83",
      year           = "2011",
      pages          = "065206",
      doi            = "10.1103/PhysRevC.83.065206",
      eprint         = "1102.4376",
      archivePrefix  = "arXiv",
      primaryClass   = "nucl-th",
      SLACcitation   = "%%CITATION = ARXIV:1102.4376;%%"
}

@article{Roberts:2011cf,
      author         = "Roberts, Hannes L. L. and Chang, Lei and Cloet, Ian C.
                        and Roberts, Craig D.",
      title          = "{Masses of ground and excited-state hadrons}",
      journal        = "Few Body Syst.",
      volume         = "51",
      year           = "2011",
      pages          = "1-25",
      doi            = "10.1007/s00601-011-0225-x",
      eprint         = "1101.4244",
      archivePrefix  = "arXiv",
      primaryClass   = "nucl-th",
      SLACcitation   = "%%CITATION = ARXIV:1101.4244;%%"
}

@article{Boucaud:2011ug,
      author         = "Boucaud, Ph. and Leroy, J. P. and Yaouanc, A. Le and
                        Micheli, J. and Pene, O. and Rodriguez-Quintero, J.",
      title          = "{The Infrared Behaviour of the Pure Yang-Mills Green
                        Functions}",
      journal        = "Few Body Syst.",
      volume         = "53",
      year           = "2012",
      pages          = "387-436",
      doi            = "10.1007/s00601-011-0301-2",
      eprint         = "1109.1936",
      archivePrefix  = "arXiv",
      primaryClass   = "hep-ph",
      SLACcitation   = "%%CITATION = ARXIV:1109.1936;%%"
}

@article{Ebert:1996vx,
      author         = "Ebert, Dietmar and Feldmann, Thorsten and Reinhardt,
                        Hugo",
      title          = "{Extended NJL model for light and heavy mesons without q
                        - anti-q thresholds}",
      journal        = "Phys. Lett.",
      volume         = "B388",
      year           = "1996",
      pages          = "154-160",
      doi            = "10.1016/0370-2693(96)01158-6",
      eprint         = "hep-ph/9608223",
      archivePrefix  = "arXiv",
      primaryClass   = "hep-ph",
      reportNumber   = "DESY-96-214, HUB-EP-96-40",
      SLACcitation   = "%%CITATION = HEP-PH/9608223;%%"
}

@article{Roberts:2007jh,
      author         = "Roberts, C. D. and Bhagwat, M. S. and Holl, A. and
                        Wright, S. V.",
      title          = "{Aspects of hadron physics}",
      journal        = "Eur. Phys. J. ST",
      volume         = "140",
      year           = "2007",
      pages          = "53-116",
      doi            = "10.1140/epjst/e2007-00003-5",
      eprint         = "0802.0217",
      archivePrefix  = "arXiv",
      primaryClass   = "nucl-th",
      SLACcitation   = "%%CITATION = ARXIV:0802.0217;%%"
}

@article{Chen:2026nf,
  author  = {Chen, Yi-huai and others},
  title   = {Flavor dependence of chiral symmetry breaking 
             and the conformal window},
  journal = {arXiv:2602.14489},
  year    = {2026},
  eprint  = {2602.14489},
  archivePrefix = {arXiv}
}

@article{Roberts:1994hh,
  author  = {Roberts, Craig D. and Williams, Anthony G.},
  title   = {Dyson-Schwinger equations and their application 
             to hadronic physics},
  journal = {Prog. Part. Nucl. Phys.},
  volume  = {33},
  pages   = {477},
  year    = {1994},
  doi     = {10.1016/0146-6410(94)90049-3}
}

@article{Munczek:1994zz,
  author  = {Munczek, H. J.},
  title   = {Dynamical chiral symmetry breaking, Goldstone's theorem 
             and the consistency of the Schwinger-Dyson and 
             Bethe-Salpeter equations},
  journal = {Phys. Rev. D},
  volume  = {52},
  pages   = {4736},
  year    = {1995},
  doi     = {10.1103/PhysRevD.52.4736}
}

@article{Maris:2000sk,
  author  = {Maris, Pieter and Tandy, Peter C.},
  title   = {The $\pi$, $K^+$, and $K^0$ electromagnetic 
             form factors},
  journal = {Phys. Rev. C},
  volume  = {62},
  pages   = {055204},
  year    = {2000},
  eprint  = {nucl-th/0005015},
  archivePrefix = {arXiv},
  doi     = {10.1103/PhysRevC.62.055204}
}

@article{Maris:1999nt,
  author  = {Maris, Pieter and Tandy, Peter C.},
  title   = {The quark photon vertex and the pion charge radius},
  journal = {Phys. Rev. C},
  volume  = {61},
  pages   = {045202},
  year    = {2000},
  eprint  = {nucl-th/9905056},
  archivePrefix = {arXiv},
  doi     = {10.1103/PhysRevC.61.045202}
}

@article{Roberts:2021nhw,
  author  = {Roberts, Craig D. and Richards, David G. 
             and Horn, Tanja and Chang, Lei},
  title   = {Insights into the emergence of mass from 
             studies of pion and kaon structure},
  journal = {Prog. Part. Nucl. Phys.},
  volume  = {120},
  pages   = {103883},
  year    = {2021},
  eprint  = {2102.01579},
  archivePrefix = {arXiv},
  doi     = {10.1016/j.ppnp.2021.103883}
}

@article{Horn:2016rip,
  author  = {Horn, Tanja and Roberts, Craig D.},
  title   = {The pion: an enigma within the Standard Model},
  journal = {J. Phys. G},
  volume  = {43},
  pages   = {073001},
  year    = {2016},
  eprint  = {1602.04016},
  archivePrefix = {arXiv},
  doi     = {10.1088/0954-3899/43/7/073001}
}

@article{Amendolia:1986wj,
  author  = {Amendolia, S. R. and others},
  collaboration = {NA7},
  title   = {A measurement of the space-like pion 
             electromagnetic form factor},
  journal = {Nucl. Phys. B},
  volume  = {277},
  pages   = {168},
  year    = {1986},
  doi     = {10.1016/0550-3213(86)90437-2}
}

@article{Amendolia:1984nz,
    author = "Amendolia, S. R. and others",
    title = "{A Measurement of the Pion Charge Radius}",
    reportNumber = "CERN-EP/84-59",
    doi = "10.1016/0370-2693(84)90655-5",
    journal = "Phys. Lett. B",
    volume = "146",
    pages = "116--120",
    year = "1984"
}

@article{Amendolia:1986ui,
    author = "Amendolia, S. R. and others",
    title = "{A Measurement of the Kaon Charge Radius}",
    reportNumber = "CERN-EP/86-98",
    doi = "10.1016/0370-2693(86)91407-3",
    journal = "Phys. Lett. B",
    volume = "178",
    pages = "435--440",
    year = "1986"
}

@article{Horn:2006tm,
  author  = {Horn, T. and others},
  collaboration = {Jefferson Lab F($\pi$)-2},
  title   = {Determination of the Charged Pion Form Factor 
             at $Q^2 = 1.60$ and $2.45$~{GeV}$^2$},
  journal = {Phys. Rev. Lett.},
  volume  = {97},
  pages   = {192001},
  year    = {2006},
  eprint  = {nucl-ex/0607005},
  archivePrefix = {arXiv},
  doi     = {10.1103/PhysRevLett.97.192001}
}

@article{Huber:2008id,
  author  = {Huber, G. M. and others},
  collaboration = {Jefferson Lab F($\pi$)},
  title   = {Charged pion form factor between 
             $Q^2 = 0.60$ and $2.45$~{GeV}$^2$},
  journal = {Phys. Rev. C},
  volume  = {78},
  pages   = {045203},
  year    = {2008},
  eprint  = {0809.3052},
  archivePrefix = {arXiv},
  doi     = {10.1103/PhysRevC.78.045203}
}

@article{AbdulKhalek:2021gbh,
  author  = {Abdul Khalek, R. and others},
  title   = {Science requirements and detector concepts for the 
             electron-ion collider: {EIC} yellow report},
  journal = {Nucl. Phys. A},
  volume  = {1026},
  pages   = {122447},
  year    = {2022},
  eprint  = {2103.05419},
  archivePrefix = {arXiv},
  doi     = {10.1016/j.nuclphysa.2022.122447}
}

@article{Dally:1980dj,
    author = "Dally, E. B. and others",
    title = "{DIRECT MEASUREMENT OF THE NEGATIVE KAON FORM-FACTOR}",
    reportNumber = "FERMILAB-PUB-80-117-E",
    doi = "10.1103/PhysRevLett.45.232",
    journal = "Phys. Rev. Lett.",
    volume = "45",
    pages = "232--235",
    year = "1980"
}

@article{Dally:1982zk,
  title = {Elastic-Scattering Measurement of the Negative-Pion Radius},
  author = {Dally, E. B. and Hauptman, J. M. and Kubic, J. and Stork, D. H. and Watson, A. B. and Guzik, Z. and Nigmanov, T. S. and Riabtsoy, V. D. and Tsyganov, E. N. and Vodopianov, A. S. and Beretvas, A. and Grigorian, A. and Tompkins, J. C. and Toohig, T. E. and Wehmann, A. A. and Poirier, J. A. and Rey, C. A. and Volk, J. T. and Rapp, P. D. and Shepard, P. F.},
  journal = {Phys. Rev. Lett.},
  volume = {48},
  issue = {6},
  pages = {375--378},
  numpages = {0},
  year = {1982},
  month = {Feb},
  publisher = {American Physical Society},
  doi = {10.1103/PhysRevLett.48.375},
  url = {https://link.aps.org/doi/10.1103/PhysRevLett.48.375}
}

@article{Gasser:1983yg,
    author = "Gasser, J. and Leutwyler, H.",
    title = "{Chiral Perturbation Theory to One Loop}",
    reportNumber = "CERN-TH-3689",
    doi = "10.1016/0003-4916(84)90242-2",
    journal = "Annals Phys.",
    volume = "158",
    pages = "142",
    year = "1984"
}

@article{Barkov:1985ac,
    author        = "Barkov, L. M. and others",
    collaboration = "OLYA, CMD",
    title         = "{Electromagnetic Pion Form-Factor in the Timelike Region}",
    journal       = "Nucl. Phys. B",
    volume        = "256",
    pages         = "365--384",
    year          = "1985",
    doi           = "10.1016/0550-3213(85)90399-2"
}

@article{Nambu:1961tp,
    author = "Nambu, Y. and Jona-Lasinio, G.",
    title = "{Dynamical Model of Elementary Particles Based on an Analogy with Superconductivity. I}",
    journal = "Phys. Rev.",
    volume = "122",
    pages = "345--358",
    year = "1961"
}

@article{Nambu:1961fr,
    author = {Nambu, Y. and Jona-Lasinio, G.},
    title = {{Dynamical Model of Elementary Particles Based on an Analogy with Superconductivity. II}},
    journal = {Phys. Rev.},
    volume = {124},
    pages = {246--254},
    year = {1961},
    doi = {10.1103/PhysRev.124.246},
    url = {https://link.aps.org/doi/10.1103/PhysRev.124.246}
}

@article{Bernard:1995hm,
title = {Pion Observables in the Extended NJL Model with Vector and Axial-Vector Mesons},
journal = {Annals of Physics},
volume = {249},
number = {2},
pages = {499-531},
year = {1996},
issn = {0003-4916},
doi = {https://doi.org/10.1006/aphy.1996.0081},
url = {https://www.sciencedirect.com/science/article/pii/S0003491696900810},
author = {Véronique Bernard and Alex H. Blin and Brigitte Hiller and Yuri P. Ivanov and Alexander A. Osipov and Ulf-G. Meißner}
}

@article{Bijnens:1995ww,
title = {Chiral Lagrangians and Nambu-Jona-Lasinio like models},
journal = {Physics Reports},
volume = {265},
number = {6},
pages = {370-446},
year = {1996},
issn = {0370-1573},
doi = {https://doi.org/10.1016/0370-1573(95)00051-8},
url = {https://www.sciencedirect.com/science/article/pii/0370157395000518},
author = {Johan Bijnens}
}

@article{Gasser:1984gg,
    author = "Gasser, J. and Leutwyler, H.",
    title = "{Chiral Perturbation Theory: Expansions in the Mass of the Strange Quark}",
    doi = "10.1016/0550-3213(85)90492-4",
    journal = "Nucl. Phys. B",
    volume = "250",
    pages = "465--516",
    year = "1985"
}

@article{Gasser:1984ux,
    author = "Gasser, J. and Leutwyler, H.",
    title = "{Low-Energy Expansion of Meson Form-Factors}",
    doi = "10.1016/0550-3213(85)90493-6",
    journal = "Nucl. Phys. B",
    volume = "250",
    pages = "517--538",
    year = "1985"
}

@article{ParticleDataGroup:2024cfk,
    author = "Navas, S. and others",
    collaboration = "Particle Data Group",
    title = "{Review of particle physics}",
    doi = "10.1103/PhysRevD.110.030001",
    journal = "Phys. Rev. D",
    volume = "110",
    number = "3",
    pages = "030001",
    year = "2024"
}

@article{McLerran:2007qj,
    author = "McLerran, Larry and Pisarski, Robert D.",
    title = "{Phases of cold, dense quarks at large N(c)}",
    eprint = "0706.2191",
    archivePrefix = "arXiv",
    primaryClass = "hep-ph",
    doi = "10.1016/j.nuclphysa.2007.08.013",
    journal = "Nucl. Phys. A",
    volume = "796",
    pages = "83--100",
    year = "2007"
}

\end{document}